\documentclass[letterpaper]{article}
\usepackage{natbib,alifexi}
\newcommand{\be}{\begin{eqnarray}}
\newcommand{\ee}{\end{eqnarray}}
\usepackage{graphicx}

\title{Critical properties of complex fitness landscapes}
\author{Bj{\o}rn {\O}stman$^{1}$, Arend Hintze$^{1}$, \and Christoph Adami$^{1}$\\
\mbox{}\\
$^1$Keck Graduate Institute, Claremont, CA 91711 \\
bostman@kgi.edu}

\begin{document}
\maketitle

\begin{abstract}
Evolutionary adaptation is the process that increases the fit of a population to the fitness landscape it inhabits. As a consequence,
evolutionary dynamics is shaped, constrained, and channeled, by that fitness landscape. Much work has been expended to understand the
evolutionary dynamics of adapting populations, but much less is known about the structure of the landscapes.
Here, we study the global and local structure of complex fitness landscapes of interacting loci that describe protein folds or sets of
interacting genes forming pathways or modules. We find that in these landscapes, high peaks are more likely to be found near other high peaks, corroborating Kauffman's ``Massif Central" hypothesis. We study the clusters of peaks as a function of the ruggedness of the landscape and find that this clustering allows peaks to form interconnected networks. These networks undergo a percolation phase transition as a function of minimum peak height, which indicates that evolutionary trajectories that take no more than two mutations to shift from peak to peak can span the entire genetic space. These networks have implications for evolution in rugged landscapes, allowing adaptation to proceed after a local fitness peak has been ascended.
\end{abstract}

\section{Introduction}

The structure of the fitness landscapes that populations find themselves in determines to a large extent how those populations will evolve. In introducing the concept of an adaptive fitness landscape, Sewall Wright~(\citeyear{Wright1932}) sought to illustrate the idea that some combinations of characters will give rise to very high fitness (peaks) while some others do not (valleys), and to study the processes that allow a population to shift from peak to peak.  Evolution in simple smooth landscapes (where each site or locus contributes independently to fitness) is trivial, because the ascent of a single fitness peak is largely deterministic~\citep{Tsimringetal1996,Kessleretal1997}.  At the other extreme lie ``random" landscapes~\citep{DerridaPeliti1991, FlyvbjergLautrup1992}, which are characterized by an absence of any fitness correlations between genotypes, and whose dynamics can likewise be solved using statistical approaches. In between these two extremes lie fitness landscapes that are neither smooth nor random, where mutations at different loci interact in complex patterns, giving rise to variedly rugged and highly {\em epistatic} landscapes~\citep{Whitlock1995,BurchChao1999,Phillipsetal2000,Beerenwinkel2007,Phillips2008}. Experiments with bacteria and viruses~\citep{ElenaLenski2003} have revealed that real fitness landscapes are of this nature: they are neither smooth nor random, and consist of a large number of fitness peaks.

Unfortunately, while experiments with bacteria and viruses have taught us a lot about evolutionary dynamics, they can only probe very limited regions of the fitness landscape, confined to the genotype space surrounding those of living organisms. In artificial landscapes we are not constrained by generation time or the specific genotypic space that organisms happen to occupy, but can place organisms anywhere in the fitness landscape, thus enabling us to examine the statistical properties of fitness landscapes.

If realistic fitness landscapes are neither smooth (a single peak) nor random (very many randomly placed peaks in the landscape), what is the structure of complex landscapes in ``peak space"? Are most peaks confined to one region of genotype space, leaving other areas empty? Are peaks clustered or are they evenly distributed? One hypothesis about the structure of fitness landscapes was proposed by Kauffman~(\citeyear{Kauffman1993}), who posited that peaks are not evenly distributed, but that high peaks are correlated in space, forming a \emph{Massif Central}, and presented numerical evidence supporting this view. According to this observation, the best place to look for a high fitness peak is near another high fitness peak. A corollary to this hypothesis is that large basins with no peaks surrounds the central massif. If fitness peaks are indeed distributed in this manner, it would have profound implications for the {\em traversability} of the landscape, and for evolvability in general~\citep{AltenbergWagner1996}.

Here we strive to study this question in much more detail, by analyzing all the peaks in a landscape in which the ruggedness can be tuned from smooth to random. In particular, we would like to know whether the highest peaks form clusters of connected walks that can {\em percolate}, i.e., form connected clusters that span the entire fitness landscape. Such clusters are very different from the neutral networks studied elsewhere~\citep{Nimwegen1999,Wilke2001}, and we briefly argue that peak networks may be more important for evolvability.

\section{NK Landscape}
\begin{figure}[htp]
\begin{center}
\includegraphics[width=2.5in,angle=0]{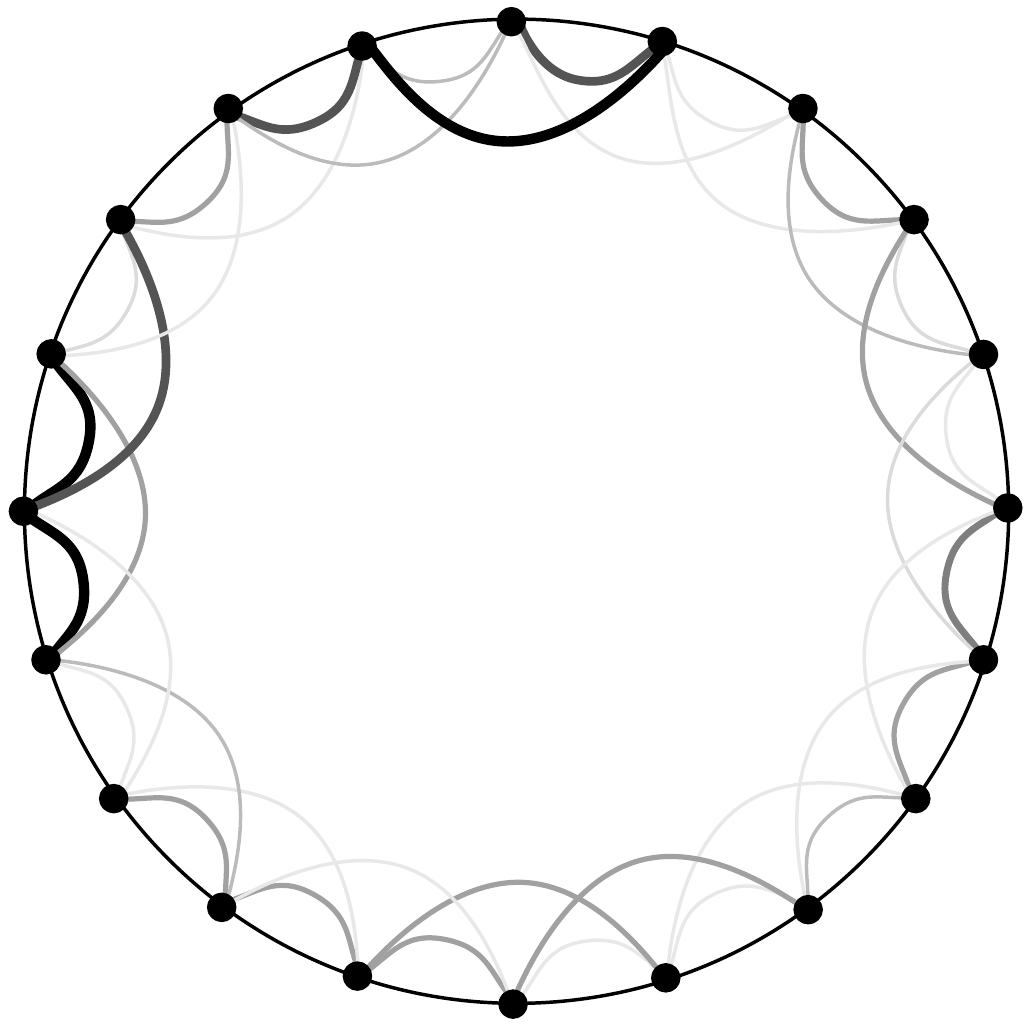}
\caption{Genome and epistatic interactions between sites for the peak genotype of an $N=20$ and $K=2$ model. While all sites within a ``radius" of two interact (light grey), the strength of interaction can be very different depending on the actual landscape that was formed. Here, the strength of epistatic interactions was calculated by performing all single-site and pairwise knockouts on the global peak genotype, and calculating the deviation of independence using a standard method~\citep{Bonhoefferetal2004,ElenaLenski1997,Ostmanetal2010}.}
\label{circ}
\end{center}
\end{figure}

Kauffman's NK model~(\citealp{KauffmanLevin1987}, see also \citealp{Altenberg1997}) has been used extensively to study evolution because it is a computationally tractable model of $N$ binary interacting loci where the ruggedness of the landscape can be tuned by varying $K$, the number of loci that each locus interacts with. Typically $N$ is of the order of 10-30, but larger sets can be studied if a complete enumeration of genotypes is not necessary. If $K=0$, the smooth landscape limit is reached, because if loci do not interact, then there is a single peak in the landscape that can be reached by optimizing each locus independently. If $K=N-1$, on the other hand, the model reproduces the random energy model of Derrida~\citep{DerridaPeliti1991}. The $N$ loci are usually thought of as occupying sites on a circular genome, while the interactions occur between {\em adjacent} sites (see Fig.~\ref{circ}), but the identity of the interactors are immaterial and the results do not depend on their physical location on the genome. The example genome in Fig.~\ref{circ} shows the interactions between loci in an $N=20$ and $K=2$ model, where the width and darkness of the lines reflects the strength of the epistatic interactions between sites for the global peak of that landscape.

While clearly the NK model should not be thought of as describing the genome of whole organisms, the model has been used extensively to study the evolution of a smaller set of sites, such as the residues in a protein~\citep{MackenPerelson1989,PerelsonMacken1995,Hayashietal2006,CarneiroHartl2010} or the set of interacting genes coding for a pathway or a module~\citep{KauffmanWeinberger1989,Soleetal2003,Yukilevichetal2008,Ostmanetal2010}.

In the original NK model, the fitness contribution of each locus is calculated as the arithmetic mean of the fitness contributions of each locus $w(x_i)$, which itself is a function of the value of the bit at that locus ('1' if the gene is expressed, '0' if it is silent) and the allele of the $K$ genes it interacts with. This fitness landscape is constructed by obtaining uniformly distributed independent random numbers for all the possible combinations of the $K+1$ sites ($2^{K+1}$ numbers for each locus), so that the fitness contribution for any combinations of alleles can simply be found by looking up that value in the table. Here, we modify this model slightly, by replacing the customary arithmetic mean by the geometric one, so that the fitness of genotype $\vec x=(x_1,...,x_N)$ is given by
\be
W(\vec x)=\left(\prod_{i=1}^Nw(x_i)\right)^{1/N}\;.
\ee
This modification better captures the nature of real genetic interactions (see, e.g.,~\citealp{StOngeetal2007}), and it makes it possible to introduce lethal mutations by setting one or more numbers in the fitness lookup-table to zero. Taking the geometric mean skews the distribution of genotype fitness to the left, resulting in a mean of about $0.4$, rather than the value of $0.5$ when using the arithmetic mean (see Fig.~\ref{fig1}). Of course the logarithm of $W(\vec x)$ reduces to the usual arithmetic mean of the log-transformed fitnesses.

In the NK model we can easily compute the fitness of all genotypes as long as $N$ and $K$ are not too large, and we can also identify {\em fitness peaks} as those genotypes whose $N$ one-mutation neighbors all have lower fitness. Increasing $K$ creates landscapes that are increasingly rugged, containing more and higher peaks with deeper valleys in between. The waiting time to new mutations becomes a determining factor in how much the population can evolve before it risks becoming stuck on a peak of suboptimal fitness. Visualizing natural fitness landscapes is difficult since it requires probing genotype-space by measuring the fitness of organisms whose genomes are fully sequenced. Even worse, natural fitness landscapes are rarely static, making such an endeavor even more futile. In computational models all genotypes can sometimes be enumerated, and we can thus learn about the global properties of the fitness landscape. This exciting possibility is muted by the fact that we cannot easily visualize high-dimensional spaces, and we are forced to resorting to statistical methods to probe the landscape.

\section{How Peaks Cluster}
\begin{figure}[htp]
\begin{center}
\includegraphics[width=3.3in,angle=0]{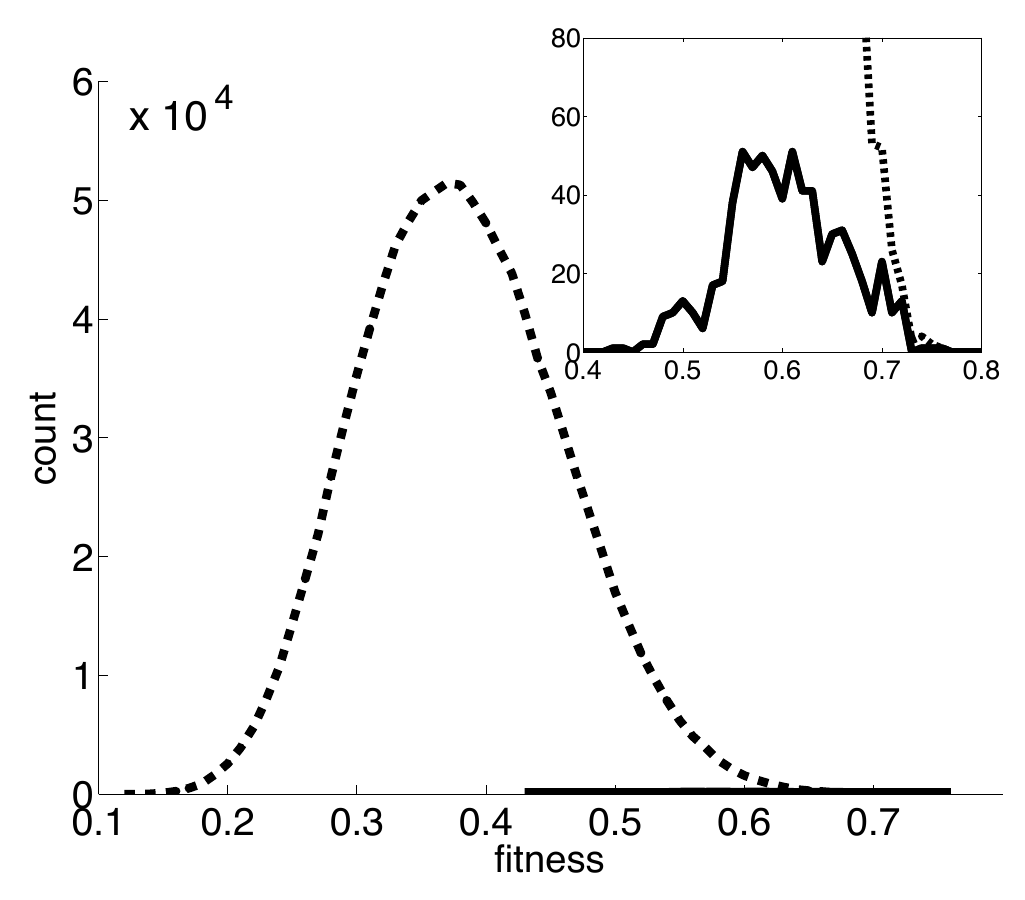}
\caption{Fitness distribution of all $1,048,576$ genotypes (dashed line) in a typical landscape of $N=20$ and $K=4$. This landscape contains 679 peaks whose fitness distribution is shown as a solid black line. In the inset we have zoomed in on the peaks.}
\label{fig1}
\end{center}
\end{figure}

In Fig.~\ref{fig1} we show the fitness distribution of all genotypes of an $N=20,K=4$ landscape (this distribution is virtually identical for different realizations of landscapes with the same $N$ and $K$). Of those $2^{20}$ genotypes, less than 0.07\% are peaks (this fraction depends on the particular realization of the landscape), and are also roughly normally distributed in fitness. Note that while the highest-fitness genotypes are very likely peaks, there are peaks whose fitness is significantly smaller, down to the mean fitness of genotypes in the landscape. The number of peaks scales approximately exponentially with $N$ (when $K$ is fixed), but only about linearly with $K$ for $K$ sufficiently large, and at fixed $N$ (data not shown).

\subsection{Pairwise distances}
Because the ``Massif Central" hypothesis says that the neighborhoods of high peaks are the best places to look for other high peaks, it is natural to also look at the pairwise distance of all peaks in a landscape. As we now know the genotypes of all the peaks in the landscape, we can ask whether peaks have a tendency to be located close to each other by studying the distribution of {\em Hamming distances} between peaks, which counts the number of differences in the binary representation of the sequences.
In fact, this is how Kauffman validated his hypothesis: by plotting the fitness of peaks as a function of the Hamming distance of all peaks to the highest peak he found~(\cite{Kauffman1993}, page 61), for a landscape with $N=96$ and $K=2$, $4$, and $8$. As it is not possible to enumerate $2^{96}\approx 8 \cdot 10^{28}$ genotypes, Kauffman found high peaks using random uphill walks. Here, we instead use $N=20$, for which we can compute the fitness of all genotypes and thus locate all peaks. After computing the Hamming distance between all pairs of peaks, we can compare the distribution of these distances to a control distribution constructed with the same number of {\em random} genotypes, which are not expected to show any bias in the distribution of distances. (It is easy to see that the distribution of pairwise distances of random binary sequences of length $N=20$ peaks at $d=10$.)

\begin{figure}[thp]
\begin{center}
\includegraphics[width=3.3in,angle=0]{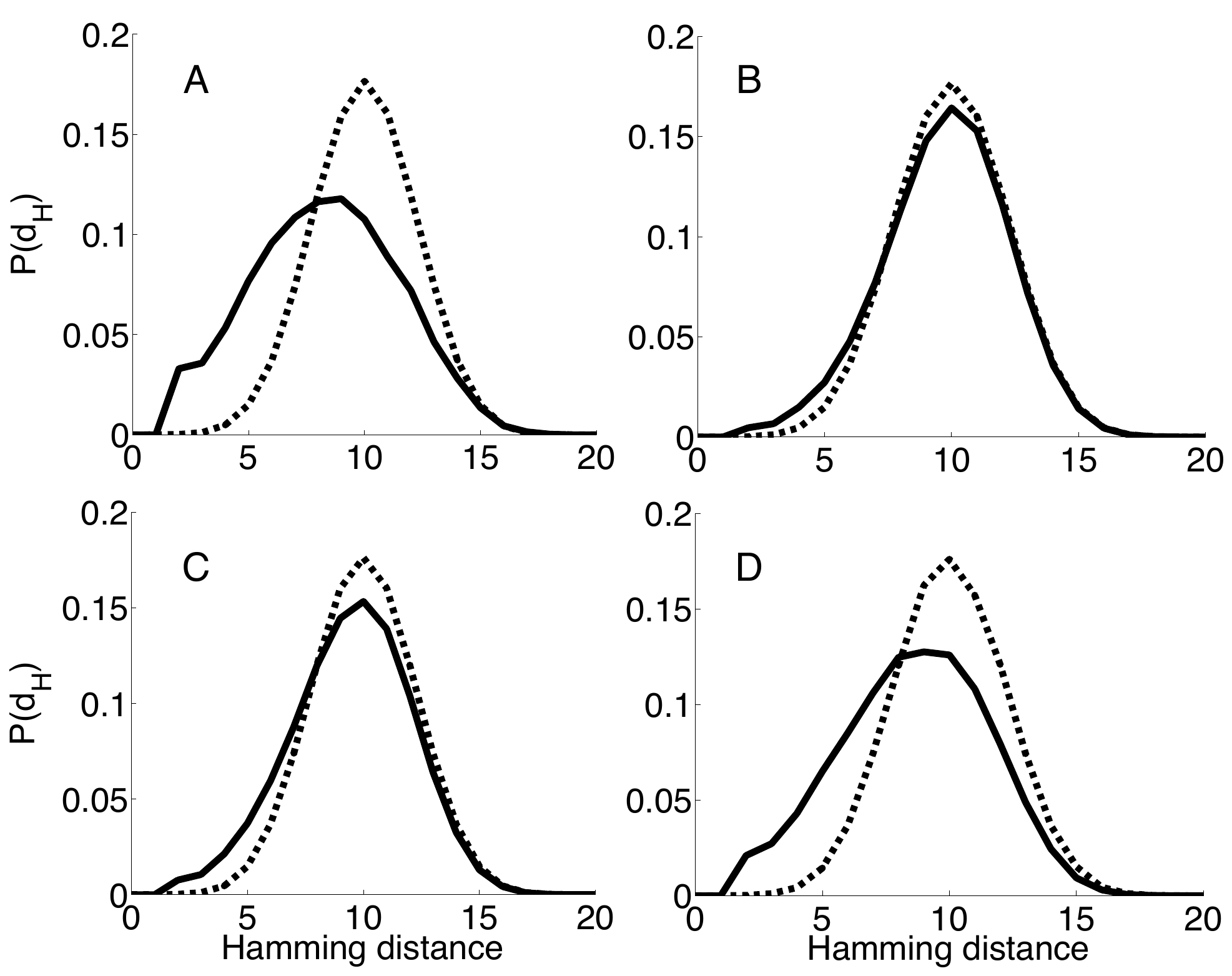}
\caption{Distributions of pairwise Hamming distances between all peaks (solid) and between random ``control" genotypes (dashed). The distributions shown are the averages of 50 different landscapes with genomes of length $N=20$. (A) $K=2$ landscapes containing an average of 98 peaks. (B) $K=4$ landscapes containing an average of $720$ peaks. (C) $K=4$ landscapes including only an average of $363$ peaks with a fitness above a threshold: $W\geq\Theta=0.60$. (D) $K=4$ landscapes including only an average of 95 peaks with a fitness above a threshold of $\Theta=0.66$. As the samples include fewer and higher peaks, the pairwise distributions of $K=4$ landscapes begin to resemble that of the $K=2$ landscapes, suggesting that the highest peaks do cluster in genotype space, whereas the distribution of lower peaks is less biased.}
\label{fig2}
\end{center}
\end{figure}

We find that for $K=2$, peaks are generally closer to each other than expected, indicating that peaks cluster in genotype space (see Fig.~\ref{fig2}A). This alone does not tell us whether high peaks are more frequently associated with other high peaks (as opposed to peaks of lower fitness). Moreover, when examining $K=4$ landscapes (that contain over seven times as many peaks on average as for $K=2$) we notice that the tendency for peaks to cluster close to each other is nearly gone, that is, the distribution closely resembles the random control (Fig.~\ref{fig2}B). However, the bias reappears when we filter the peaks so that we only include those of high fitness (Figs.~\ref{fig2}C and D), reaffirming the hypothesis that in complex epistatic landscapes,  there is something special about being a high peak, genotypically speaking.

\begin{figure}[htp]
\begin{center}
\includegraphics[width=3.4in,angle=0]{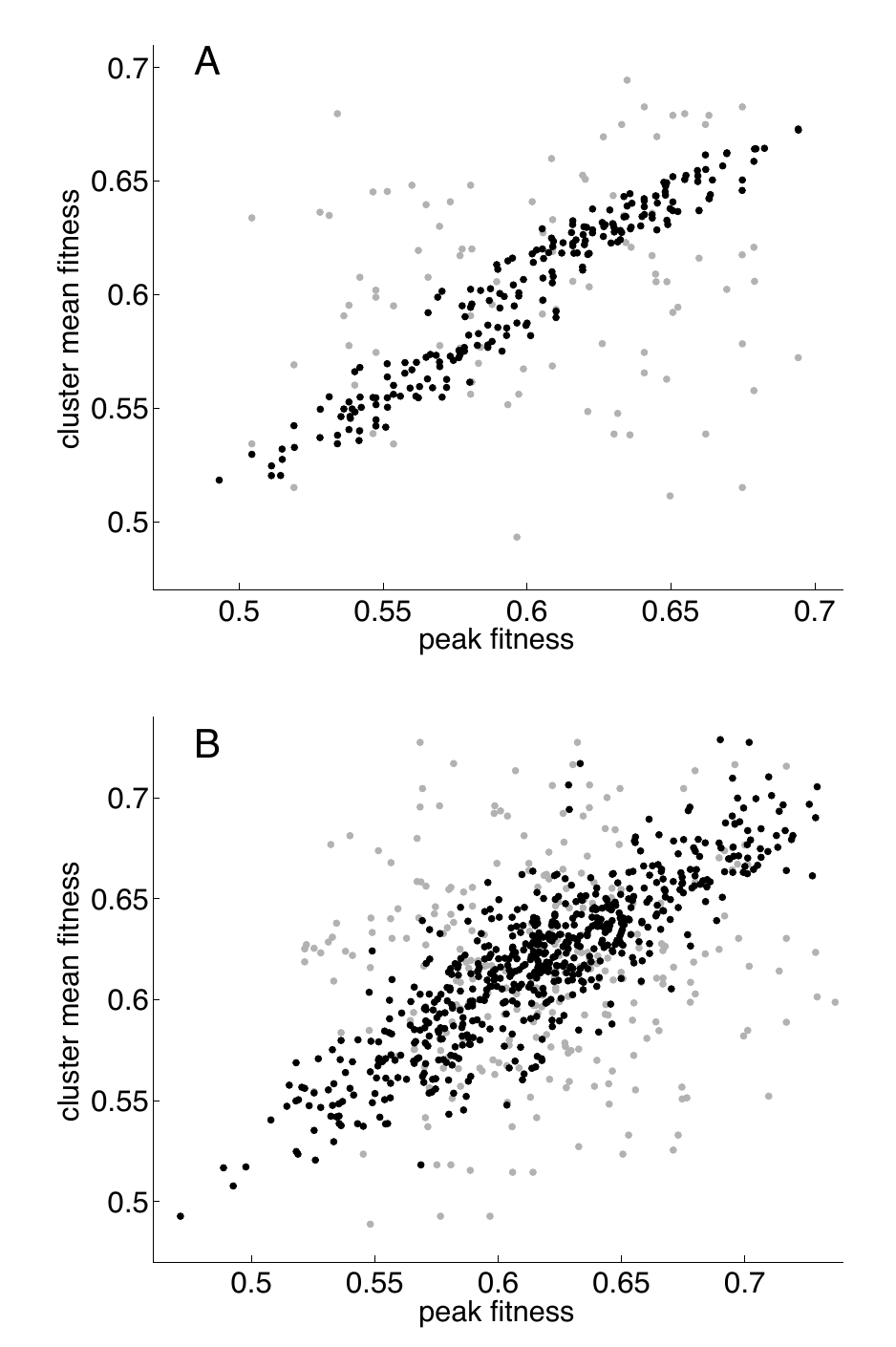}
\caption{Mean fitness of peaks in circular clusters of radius $d=2$ as a function of the fitness of the peak in the center of the cluster. (A) One landscape of $K=2$ with $166$ peaks (black dots). All landscapes show a strong correlation between cluster mean fitness and peak fitness, while the same analysis of assigning random genotypes to the peaks (but keeping the fitness) shows no such correlation (gray dots). The random data are from ten samplings. (B) One landscape of $K=4$ with $679$ peaks (black dots), and random genotypes (gray dots) obtained by sampling four times.}
\label{fig3}
\end{center}
\end{figure}

\subsection{Peak neighborhood}
If we want to know whether peaks with high fitness are likely to be found near other such peaks, we should study the mean fitness of peaks within a specified radius of that peak.  These ``circular" clusters contain all peaks within a Hamming distance $d$ of a chosen peak (not counting the peak at the center). For the smallest possible distance between peaks $d=2$, the size of a cluster is limited to $210$ genotypes, but since peaks must be at least two mutations away from each other, there can be at most $190$ peaks within a Hamming distance of two.

Fig.~\ref{fig3}A depicts the mean fitness of adjacent peaks in circular clusters of radius $d=2$ (black dots, for $K=2$), showing a tight correlation between peak fitness and average adjacent peak fitness that indicates that the immediate neighborhood of high peaks is populated by other peaks of high fitness. On the contrary, when we randomize the location of the 166 peaks in genotype space without changing their height, this relationship vanishes (light gray dots in Fig.~\ref{fig3}A). For $K=2$ random peaks are far apart, resulting in only very few peaks within a distance $d=2$ of each other. The $K=4$ landscape has four times as many peaks as the $K=2$ landscape, and the effect persists (Fig.~\ref{fig3}B). The observed relation between mean fitness of these circular clusters and peak fitness persists even when the radius in increased to $d=6$ (data not shown). We observe a similar correlation between mean cluster fitness and maximum peak height in network clusters (data not shown).

\subsection{Adjacency matrices}
\begin{figure}[htp]
\begin{center}
\includegraphics[width=3.3in,angle=0]{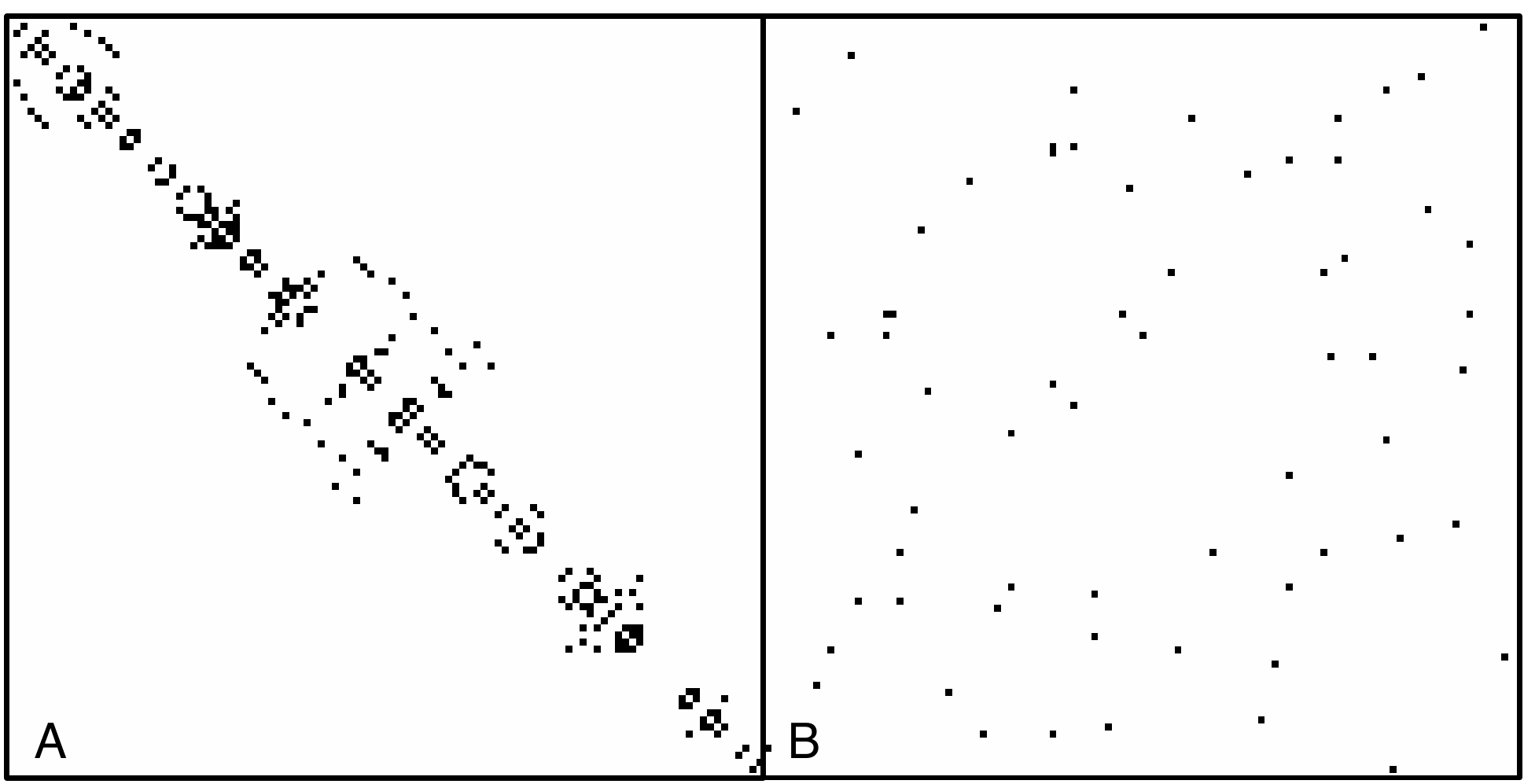}
\caption{Adjacency matrices showing clusters of peaks. (A) Single $K=4$ landscape with peaks of Hamming distance $d=2$ connected. The peaks are ordered according to which network cluster they belong to. This landscape consists of $109$ peaks with fitness above $\Theta=0.66$ that are grouped into nine clusters (not counting singletons). (B) Random $K=4$ landscape with $d=4$ and $\Theta=0$, showing only the first $109$ genotypes.}
\label{fig4}
\end{center}
\end{figure}

While circular clusters can tell us whether high peaks are surrounded by peaks that are higher than expected, they do not allow us to examine certain critical properties of the landscape. To do this, we should think of peaks in the genetic landscape as nodes in a random graph, and study the size of clusters of peaks that are formed by connecting all those peaks that are within a distance $d$ of each other. Connecting such \emph{networks clusters} of peaks creates a {\em percolation problem} (see, e.g.,~\cite{BollobasRiordan2006}). In statistical physics, systems where nodes are connected by edges that are placed with a fixed probability undergo a geometric phase transition as a function of the edge placement probability. One of the quantities studied in percolation theory is the size of the largest cluster, because this variable rises dramatically at the critical point so that it takes up most of the system once past the critical point. If the largest cluster takes up most of the nodes, the system is said to "percolate", which implies that the cluster spans the entire system (allowing you to walk across connected nodes from any part to any other in the system). We will study the percolation properties of the fitness landscape by using the peak height as the critical parameter. Clearly, if only the highest few peaks are considered the system is far from percolation, as these peaks are unlikely to be connected. But if the highest peaks are closer to each other than expected in a random control, then the peaks could percolate far earlier.

Let us begin by computing the Hamming distance between all pairs of peaks with fitness greater than $\Theta$, and connect those peaks that are a distance of no more than $d$ away from each other. In Fig.~\ref{fig4}A, we show the {\em adjacency matrix} of clusters, which we obtained by placing a dot for every two peaks that are with a distance $d$ (that is, immediately adjacent). Peaks are ordered in such a way that peaks that fall into the same cluster are placed next to each other. This procedure allows us to the visualize the structure of clustered peaks in the landscape. In contrast, if the same peaks are assigned random locations in the landscape, there is no apparent structure, and clusters of peaks are on average very small (Fig.~\ref{fig4}B). For $K=4$ and $d=2$ very few peaks are connected in a random landscape, and because of this the adjacency matrix shown in Fig.~\ref{fig4}B is for $d=4$, and includes peaks of any height. Only the first $109$ peaks are shown.

\subsection{Percolation phase-transition}
\begin{figure}[htp]
\begin{center}
\includegraphics[width=3.3in,angle=0]{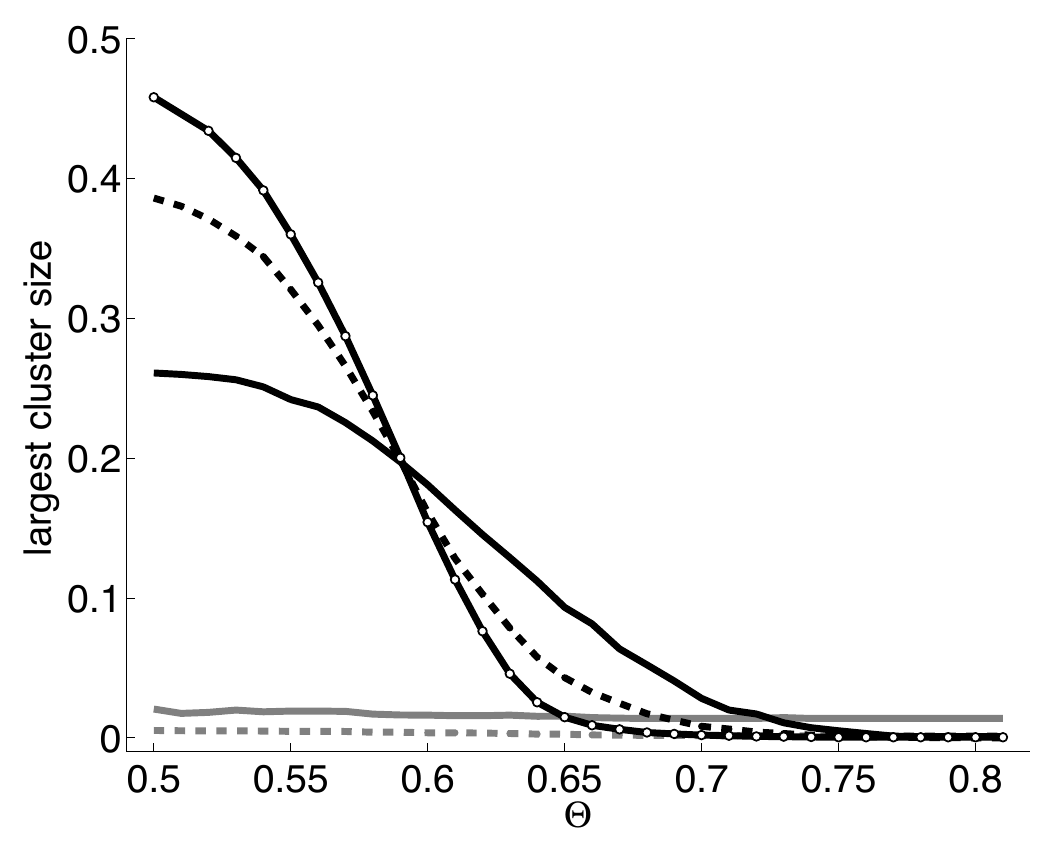}
\caption{Size of the largest network cluster in the landscape averaged over $50$ landscapes for each $K$ as a function of fitness threshold, $\Theta$. $K=2$ (solid black line), $K=4$ (dashed black line), and $K=6$ (solid black line with white circles). The more rugged the landscapes are, the more abrupt the transition is from small network clusters to one cluster dominating the landscape. Random genotypes for $K=2$ (solid gray line) and $K=4$ (dashed gray line) show no increase in cluster size.}
\label{fig5}
\end{center}
\end{figure}

In Fig.~\ref{fig5} we show the average relative size of the largest network cluster as a function of the peak  threshold $\Theta$, defined as the ratio of the largest number of connected peaks with fitness above $\Theta$ to the total number of peaks in the landscape. The relative size of the largest connected component (also called the "giant cluster" in percolation theory) increases dramatically as the critical threshold is reached, much like the size of the giant component increases when the critical probability of edges is reached in percolation theory.  But what is remarkable about this transition is that it only occurs because the high peaks in the landscape occur near other high peaks: if the peaks were not clustered, the largest network cluster size would not increase when we lower $\Theta$, as is the case when we reassign peaks to random genotypes (gray lines in Fig.~\ref{fig5}). \\

When we include enough peaks, either by setting $\Theta$ low for $K=4$ (or else for $K=6$ or higher) we find that for $d=2$ there are always {\em two} largest network clusters, while the third largest cluster contains significantly fewer peaks. Both large clusters percolate genotype space and the diameter of both graphs is 18, not 20 (in general, $N-2$), while the shortest distance between the two clusters is always 3. This is peculiar to the way clusters are formed in this particular percolation problem. It is a rewarding exercise to determine the root cause of this peculiarity, which we leave to the interested reader. The transition seen in Fig.~\ref{fig5} suggests that in more rugged landscapes there are several clusters containing high peaks (high $\Theta$), and that these high-peak clusters are connected by the peaks of lower fitness (lower $\Theta$).


The percolation of genetic space by peaks with a sufficiently low height is reminiscent of the percolation of genetic space by arbitrary shapes in the RNA folding problem~\citep{Grueneretal1996}, except that in that case structures with different genotypes form a neutral network that can be traversed by single point mutations. The giant cluster of peaks in the NK landscapes cannot be traversed like that: rather, it requires a minimum of two mutations to jump from peak to peak, and because some of the peaks have inferior fitness, such mutations can only be tolerated for a finite amount of time--long enough to jump to the next highest peak. Thus, deleterious mutations are likely to be important to reach distant areas in genotype space, and the importance of these is slowly being realized~\citep{Lenskietal2003,Lenskietal2006,Cowperthwaite2006,Ostmanetal2010}.

\section{Discussion}
Using several methods we have shown that the rugged fitness landscapes that epistatic interactions create in the NK model consist of fitness peaks that are distributed in a manner that strongly affects evolution. High peaks are more likely to be found near other high peaks, rather than near lower peaks or far from peaks altogether. Similarly, lower peaks are predominantly located near each other in genotype space. Cluster analysis reveals that peaks tend to cluster (as compared to the same peaks placed randomly in genetic space) giving rise to large basins of attraction that are effectively devoid of peaks. This feature is especially prominent for moderately rugged landscapes ($K=2$), while the addition of many more smaller peaks in more rugged landscapes ($K=4$ or higher) makes this trend less significant.
To the extent that we think that the NK landscape is an accurate model for real fitness landscapes of proteins and genetic pathways or modules, the discovery that these landscapes possess a remarkable structure that appears to be conducive to adaptation is highly informative about the process of evolution. Clustering of peaks makes a difference when the environment changes in a way that is unfavorable to the population, and forcing the population to adapt anew. If the landscape consists of evenly distributed peaks, then the risk of becoming stuck on a low fitness peak is high, and the population risks extinction. On the other hand, if peaks are unevenly distributed, then the ascent of one peak may not be where adaptation ends, making it possible to locate the global peak or another high fitness peak.

The more rugged a landscape is, the more peaks it contains, and the larger the space of genotypes that the largest network cluster spans. In smooth landscapes with only one or a few peaks, populations can evolve from genotypes of low fitness and move across genotype space toward high fitness. In rugged landscapes, the population always risks becoming stuck on a suboptimal peak. However, networks of closely connected peaks that percolate genotype space may still make it possible to traverse the fitness landscape jumping from peak to peak (given a sufficiently high mutation rate). If peaks are evenly distributed in genotype space, the chance to jump from peak to peak and thereby eventually locate the global peak is virtually nil. It is important, however, to remember that there are limits to the realism of the NK landscape as a model of realistic genetic or protein landscapes. For example, it is known that a significant percentage of substitutions in proteins or mutations in genetic pathways are neutral, while the NK landscape has virtually no neutrality (even though most mutations do not change the fitness significantly). Neutrality plays an important role to enhance traversability, and will facilitate the transition between peaks so that deleterious mutations are not essential for the shift from peak to peak. However, one could maintain that deleterious mutations are more promising for adaptation than neutral mutations are, because they may be what separate important phenotypes~\citep{Lenskietal2006}.

The observation that peaks form clustered networks, and that these networks percolate, implies that the risk of becoming stuck on a suboptimal peak is significantly mitigated, because all it takes is the two right mutations to locate a new peak. Thus, it appears that evolvability comes for free in complex rugged landscapes of interacting loci. We should note, however, that the reason {\em why} peaks cluster in landscapes with epistatic interactions is not immediately apparent, and is a subject of ongoing investigations.

\section{Acknowledgements}
The authors thank Nicolas Chaumont for contributing code. This work was supported by the National Science Foundation's Frontiers in Integrative Biological Research grant FIBR-0527023.
\footnotesize
\bibliographystyle{apalike}
\bibliography{NKlandscape}

\end{document}